\documentstyle[12pt]{article}

\begin{document}
\thispagestyle{empty}

\hspace*{12.6cm} UPR-0842-T

\hfill hep-th/9903243

\vspace*{2cm}

\begin{center}
{\Large Implications of decoupling effects 
for   one-loop corrected  effective  actions from
superstring  theory
}
\end{center}

\centerline{\large I.L. Buchbinder$^{*}$, M.
Cveti\v{c}$^{+}$ and A.Yu. Petrov$^{*}$}
\begin{center}
$^{*}${\small\it Department
of Theoretical Physics, Tomsk State Pedagogical University,}\\
 {\small\it 634041
Tomsk, Russia}

\vspace*{2mm}

 $^{+}${\small\it Department of Physics and Astronomy, University of
Pennsylvania,}\\ {\small\it Philadelphia, PA 19104--6396, USA}
\end{center}

\vspace*{3cm}

\begin{abstract}
We study  the  decoupling effects in  one-loop corrected
 $N=1$ supersymmetric theory with gauge
neutral chiral superfields, by calculating the one-loop corrected effective Lagrangian that involves
light
and heavy fields with the mass scale $M$, and subsequently  eliminating
 heavy fields  by their equations of motion. In addition to 
 new non-renormalizable couplings, we determine the terms 
    that grow as
 $\log M$ and  
renormalize the fields  and couplings in the effective field theory,
in accordance with the decoupling
 theorem. 
However, in a theory derived from superstring theory,
these  terms  can 
significantly modify  low energy predictions for the
effective couplings of light
fields. For example, in a class of  heterotic superstring vacua
 with an anomalous $U(1)$  the 
vacuum restabilization  introduces such  decoupling effects
which  in turn  correct  the low energy predictions
for certain couplings  by  10-50\%.

\end{abstract}
\newpage

In principle the  decoupling  effects of heavy  fields in   field
theory are  well understood.  According to  the decoupling theorem
~\cite{appcara} (for additional references see, e.g., \cite{Collins}) in
 field
theory of interacting light (with masses $m$) and heavy fields
(with masses $M$)
the heavy fields decouple;
the effective Lagrangian of the light fields can be written in terms of the
original
classical Lagrangian of   light  fields with  the
 loop effects  of heavy fields
absorbed into redefinitions of
 new light fields, masses and couplings, and the only
new terms in this effective  Lagrangian
 are non-renormalizable ones,  proportional to inverse powers of $M$ (both at tree-
 and loop-levels).

When  field theory  is  an effective description of
phenomena at certain energies,  the rescaling of the fields  and couplings
due to the heavy fields  does not affect the structure of  couplings,
since those are  {\it free parameters} whose
values are determined by experiments.  On the other hand if the
field theory is describing an effective theory of an underlying
fundamental theory, like superstring theory, where the couplings at the
string scale are calculable, the  decoupling effects of the heavy field
can be  important and can significantly  affect the low energy
predictions  for the couplings of light fields at low energies.
Therefore the quantitative study of decoupling effects
 at the  loop-level  in  effective
  supersymmetric theories is important; it  should
    improve our understanding of such effects for the
effective  Lagrangians from superstring theory and provide us with
 calculable corrections for the  low energy predictions of the theory.

Effective theories of $N=1$ supersymmetric  four-dimensional perturbative
superstring vacua
can be obtained by employing techniques
of  two-dimensional conformal field theory~\cite{Shenkeretal}.
In particular the k\"ahlerian and  the chiral (super-) potential
 can be calculated  explicitly at the tree level. (For a
 representative work on the
 subject see, e.g., \cite{bailin,GSW}, and references
 therein.)~\footnote{While the superpotential terms   calculated at the
  string tree-level  are protected from higher genus
 corrections, the k\"ahlerian  potential is not.  Such higher
  genus corrections to the
 k\"ahlerian potential could be
 significant, however, their structure has not been studied very much.}

One of the compelling motivations for a detailed study of decoupling effects
 is  the phenomenon of
vacuum restabilization~\cite{dsw}  for  a class of  four-dimensional
(quasi-realistic)   superstring vacua with an
``anomalous'' $U(1)$.  For  such  string vacua of perturbative heterotic  string
theory, the
Fayet-Iliopoulos  (FI) $D$- term is generated at
genus-one~\cite{atick}, thus triggering certain  fields to acquire vacuum
expectation values (VEV's) of order $M_{String}\sim g_{gauge}
M_{Planck}\sim 
\times 10^{17}
 \,$ GeV
along $D$- and $F$- flat directions of the effective $N=1$ supersymmetric
theory.~\footnote{On the open Type I string side these effects are
closely related to the blowing-up procedure of Type I orientifolds  and
were recently studied in \cite{CELW}.} (Here $g_{gauge}$ is the gauge
coupling and $M_{Planck}$ the Planck scale.) Due to these large
string-scale VEV's  a number of additional fields obtain large
string-scale masses. Some of them in turn couple through
(renormalizable) interactions to the remaining light fields, and thus through
decoupling effects,
affect the effective theory of  light fields
at low energies. (For the study of the effective Lagrangians and their
phenomenological implications for a  class  of
such four-dimensional string vacua see, e.g., \cite{cceel2}--\cite{Cv4}  and references therein.)

 The  tree level decoupling effects  within $N=1$ supersymmetric theories,
 were studied within superstring theory  in \cite{Burgess}. In  a related
 work \cite{katehou} it was  shown that 
the leading order corrections  of order $1\over M$ are to the
effective superpotential, however there are also important next order effects
in  the  effective superpotential \cite{cew}.
In addition, in \cite{cew} the nonrenormalizable modifications of the
K\"{a}hler potential  (also pointed out in \cite{dine}) were
systematically studied.
These tree level  decoupling effects (as triggered by, e.g., vacuum
restabilization for a class of string vacua)
 lead to new nonrenormalizable interactions which are competitive
with the nonrenormalizable  terms that are calculated
 directly in the superstring  theory.

The main purpose of this paper is to  address the one-loop decoupling effects in
$N=1$ supersymmetric theory. The main motivation for this  study is that 
such effective Lagragians arise  naturally as  subsectors
of an effective theory for a class superstring vacua
 with an anomalous $U(1)$; vacuum restabilization triggers couplings between heavy
fields with mass scale $M\sim 10^{17}$ GeV and the light (massless) fields. 
We specifically concentrate on the effects of  chiral (gauge neutral)
superfields. (We do not include supergravity effects and soft supersymmetry
breaking effects which could  be significant.)
For the sake of simplicity and to illustrate the effects clearly,
we shall consider a specific toy model with only one light and one heavy
chiral superfield and a specific renormalizable form of their interactions.
(A systematic presentation of the derivation of the results and applications to
more general models will be given elsewhere~\cite{Buchbinderetal}.)

The starting point is the  effective theory with   (gauge neutral) $N=1$
chiral superfields $\Phi^i$  whose  action is of the form:
\begin{equation}
\label{acti}
S[\Phi,\bar{\Phi}]=\int d^8 z K(\bar{\Phi}^i,\Phi^i)+
(\int d^6 z W(\Phi^i)+h.c.)\ , 
\end{equation}
where $\Phi^i=\Phi^i(z)$, $z^A\equiv(x^a,\theta_{\alpha},
\bar{\theta}_{\dot{\alpha}})$; $a=0,1,2,3$; $\alpha=1,2$,
$\dot{\alpha}=\dot{1},
\dot{2}$, $d^8 z= d^4 x d^2\theta d^2\bar{\theta}$. Real function
$K(\bar{\Phi}^i,\Phi^i)$ is called k\"{a}hlerian potential and holomorphic
function $W(\Phi^i)$ is called chiral potential \cite{BK0}.
We refer to the action (\ref{acti}) as a general chiral
superfield
model. It is
the  most general model constructed from chiral superfields and contains no
higher
derivatives at component level.
(The model (\ref{acti}) can be  interpreted as  a subsector of an
effective theory at energy scales $\mu\ll M_{String}$, derived from an underlying fundamental theory--
superstring theory;
potentials $K(\bar{\Phi}^i,\Phi^i)$, $W(\Phi^i)$ are calculable in 
superstring theory.)

We shall investigate the effective Lagrangian  (\ref{acti})
with  the light and heavy  chiral superfields,  with interaction terms that
involve both. 
We shall
calculate the one-loop superfield effective potential
and eliminate the heavy superfields by their equations of motion
 containing one-loop
quantum corrections. Subsequently we shall study the 
 implications of the  obtained  effective action of the
light superfield, only.

Though   the  calculations can be carried out with
arbitrary functions $K(\bar{\Phi}^i,\Phi^i)$ and $W(\Phi^i)$
(again, these generalizations will be given elsewhere~\cite{Buchbinderetal}) we
restrict
ourselves for simplicity to the  so called minimal case where
\begin{equation}
K(\Phi,\bar{\Phi},\phi,\bar{\phi})=\Phi\bar{\Phi}+\phi\bar{\phi}\ , 
\end{equation}
and
\begin{equation}
W=\frac{M}{2}\Phi^2+\frac{1}{2}\lambda\Phi\phi^2+\frac{g}{3!}\phi^3\ .
\end{equation}
Here $M$ is a mass of heavy superfield $\Phi$, the light superfield $\phi$ is
assumed massless.
Hence the total classical action is of the form:
\begin{equation}
\label{act}
S=\int d^8 z (\Phi\bar{\Phi}+\phi\bar{\phi})+\big(\int d^6 z
[\frac{M}{2}\Phi^2+\frac{1}{2}\lambda\Phi\phi^2+\frac{g}{3!}\phi^3] +h.c.\big)\
.
\end{equation}

Let $\Gamma[\Phi,\bar{\Phi},\phi,\bar{\phi}]$ be the effective action of the
model under consideration. In terms of the expansion in powers of the 
 covariant derivatives the
effective action can be written as
\begin{equation}
\Gamma[\Phi,\bar{\Phi},\phi,\bar{\phi}]=
\int d^8 z K_{eff}(\Phi,\bar{\Phi},\phi,\bar{\phi}) +
(\int d^6 z W_{eff}(\Phi,\phi) + h.c.)+\ldots\ ,
\end{equation}
where the dots denote the terms that  depend on covariant derivatives.
The functions $K_{eff}$ and $W_{eff}$ are  the k\"{a}hlerian and chiral
effective potential, respectively~\cite{BK0}. One can show that quantum
corrections to $W_{eff}$ begin at  two-loops. 
\cite{BK0,West2}.   Thus  at the  one-loop level the 
contributions are restricted to the k\"hlerian potential, only:
$K_{eff}(\Phi,\bar{\Phi},\phi,\bar{\phi})=
K(\Phi,\bar{\Phi},\phi,\bar{\phi})+K^{(1)}(\Phi,\bar{\Phi},\phi,\bar{\phi})$
where  $K$   is the  classical 
 k\"{a}hlerian potential and $K^{(1)}$ is the  one-loop
quantum correction.

We follow the following (standard) strategy.
We determine the  heavy superfields
by  solving  the their  equations of motion 
in a theory with the one-loop corrected effective action. The obtained
solution for heavy superfields  is then inserted into the
 effective action. The result is   the effective action
 $S_{eff}[\phi,\bar{\phi}]$ of light fields, only~\footnote{To be more
 precise, one  should  insert the solution for the heavy superfields
 into equations of
motion for light superfields and then derive  the  effective action of light
fields which would precisely reproduce   the latter equations
of motion. We checked that in the case under
consideration the result is the same as in the case  of
directly substituting the solution for heavy
superfields into the  original one-loop corrected effective action.}.
(The same  procedure for pure classical theory, i.e.,
$K_{eff}=K$  has been employed  in \cite{cew}.)

The equations of motion for heavy superfields in the case under consideration
have the form
\begin{eqnarray}
\label{eqmh}
 -\frac{1}{4}\bar{D}^2\big(\bar{\Phi}+
\frac{\partial K^{(1)}}{\partial\Phi}\big)+
M\Phi+\frac{\lambda}{2}\phi^2=0 \label{six} \ ,
\end{eqnarray}
and the complex
conjugate equation. We solve these equations by iterations in inverse
powers of the mass
parameter $M$ by  expanding  $\Phi=\Phi_0+\Phi_1+\ldots+\Phi_n+\ldots$,
where
$\Phi_0=-\frac{\lambda\phi^2}{2M}$ and $\Phi_n\sim \frac{1}{M^n}$.
Substituting the expansion of $\Phi$ into eq. (\ref{eqmh}) one obtains
\begin{eqnarray}
\label{ee1}
& &M\Phi_{n+1}
=\frac{\bar{D}^2}{4}(\bar{\Phi}_n+
\frac{\partial K^{(1)}}{\partial \Phi}|_{\Phi=\Phi_0+\ldots+\Phi_n}-
\frac{\partial K^{(1)}}{\partial \Phi}|_{\Phi=\Phi_0+\ldots+\Phi_{n-1}})\ , 
\end{eqnarray}
and  an analogous equation for $\bar{\Phi}_{n+1}$. In principle, these 
equations allow one to find the $n$-th order corrections $\Phi_n$ in an explicit form.

We now turn to  the calculation of $K^{(1)}$. We
employ the technique for the  computation of the
superfield effective potential developed in refs.\cite{Buch1}.  The one-loop
correction $\Gamma^{(1)}$ to the effective action can be written as
\begin{eqnarray}
\label{int}
e^{i\Gamma^{(1)}}=\int D\bar{\phi}_q D\phi_q D\Phi_q D\bar{\Phi}_q
e^{iS_2[\phi,\bar{\phi},\Phi,\bar{\Phi};
\phi_q,\bar{\phi}_q,\Phi_q,\bar{\Phi}_q]}\ ,
\end{eqnarray}
where we split the original superfields into  the background ones: 
$\phi,\bar{\phi},\Phi,\bar{\Phi}$ and  the quantum  ones:
$\phi_q,\bar{\phi}_q,\Phi_q,
\bar{\Phi}_q$,  by using  the rule $\phi\to\phi+\phi_q$,
$\bar{\phi}\to\bar{\phi}+\bar{\phi}_q$, $\Phi\to\Phi+\Phi_q$,
$\bar{\Phi}\to\bar{\Phi}+\bar{\Phi}_q$. $S_2$ is a part of the  classical
action quadratic in  the quantum superfields. The path integral (\ref{int}) is
taken over the chiral superfields. It can be transformed into  a path integral
over general superfields $u,v$ (for details, see  refs.
\cite{BK0,Buch1}):
\begin{eqnarray}
S[u,v]&=&\frac{1}{2}\int d^8 z
\Big\{ v(\Box
 -\frac{1}{4}(\lambda\Phi+g\phi)\bar{D}^2-\frac{1}{4}
(\lambda\bar{\Phi}+g\bar{\phi})D^2
)v+\nonumber\\&+&
2u(-\frac{1}{4}\lambda\phi \bar{D}^2-\frac{1}{4}\lambda\bar{\phi} D^2)v+
u(\Box-\frac{1}{4}M \bar{D}^2-\frac{1}{4}M D^2)u
\Big\}\ .
\end{eqnarray}
Eq. (\ref{int}) leads to
\begin{eqnarray}
\Gamma^{(1)}&=&\frac{i}{2}Tr\log\Delta\ , \nonumber\\
\Delta&=&\Box 1_2+
\left(
\begin{array}{cc}
 -\frac{1}{4}(\lambda\Phi+g\phi)\bar{D}^2-\frac{1}{4}
(\lambda\bar{\Phi}+g\bar{\phi})D^2&
 -\frac{1}{4}\lambda\phi \bar{D}^2-\frac{1}{4}\lambda\bar{\phi} D^2\\
 -\frac{1}{4}\lambda\phi \bar{D}^2-\frac{1}{4}\lambda\bar{\phi} D^2&
 -\frac{M}{4}(\bar{D}^2+D^2)
\end{array}\right)\ .
\end{eqnarray}
Here $1_2$ is a  unit $2\times 2$ matrix,  and the operator $\Delta$ acts 
in the space of
general superfields.
In order to cancel the divergencies in the effective action (\ref{int}) we add to the
initial action (\ref{act}) the one-loop counterterm
$
S^{(1)}_{ct}=\frac{1}{16\pi^2\epsilon}(|\lambda\Phi+g\phi|^2+
2\lambda^2|\phi|^2)
$, with $\epsilon$ being  a parameter of dimensional regularization. In the
following we  further
consider  the renormalized effective potential $K_{eff}$.

Since we are interested only in the k\"{a}hlerian effective potential we can
set all the background superfields to be  constant. Then one can use
the superfield proper-time technique \cite{BK0,Buch1} and obtain the following
form of the one-loop correction to the k\"aherian potential:
\begin{eqnarray}
\label{kahl1}
& &K^{(1)}=
 -\frac{1}{32\pi^2}
\Big(
(
|\lambda\Phi+g\phi|^2 +2 \lambda^2\phi \bar{\phi}+M^2+\\&+&
\sqrt
{(|\lambda\Phi+g\phi|^2-M^2)^2+4
|\lambda^2\Phi\bar{\phi}+\lambda M\phi +\lambda g |\phi|^2)|^2
})
\times\nonumber\\&\times&
\log\frac{
(
|\lambda\Phi+g\phi|^2 +2 \lambda^2\phi \bar{\phi}+M^2+
\sqrt
{(|\lambda\Phi+g\phi|^2-M^2)^2+4
|\lambda^2\Phi\bar{\phi}+\lambda M\phi +\lambda g |\phi|^2|^2
})
}{\mu^2}
+\nonumber\\&+&
(
|\lambda\Phi+g\phi|^2 +2 \lambda^2\phi \bar{\phi}+M^2-
\sqrt
{(|\lambda\Phi+g\phi|^2-M^2)^2+4
|\lambda^2\Phi\bar{\phi}+\lambda M\phi +\lambda g |\phi|^2|^2
})
\times\nonumber\\&\times&
\log\frac{
(
|\lambda\Phi+g\phi|^2 +2 \lambda^2\phi \bar{\phi}+M^2-
\sqrt
{(|\lambda\Phi+g\phi|^2-M^2)^2+4
|\lambda^2\Phi\bar{\phi}+\lambda M\phi +\lambda g |\phi|^2|^2
})
}{\mu^2}\Big)\ .\nonumber
\end{eqnarray}
This is the focal result of the paper. 
As a result, the low-energy effective action with one-loop quantum corrections
has the form
\begin{equation}
\label{ea}
\Gamma^{(1)}=\int d^8 z \big(\phi\bar{\phi}+\Phi\bar{\Phi}+
\hbar K^{(1)}[\phi,\bar{\phi},\Phi,\bar{\Phi}]
\big) +
\big(\int d^6 z
(\frac{M}{2}\Phi^2+\frac{\lambda}{2}\Phi\phi^2+\frac{g}{3!}\phi^3)
+ h.c.\big)\ ,
\end{equation}
where $K^{(1)}$ is given by eq. (\ref{kahl1}).  In order to keep track of the
one-loop (quantum) effects, here and in the following
we display explicitly the  Planck constant $\hbar$.

We now solve the  equations of motion for heavy superfields with the
one-loop corrected
effective action obtained above.
As explained earlier, we solve  eqs. (\ref{eqmh})-(\ref{ee1}) via
an iterative
method. In the leading order,  one obtains
\begin{equation}
\label{app3}
\Phi=-\lambda\frac{\phi^2}{2M}-
\frac{\hbar\bar{D}^2}{64\pi^2 M}
\Big(\lambda g\bar{\phi}
(1+\log\frac{2g^2|\phi|^2}{\mu^2})\Big)+O(\frac{1}{M^2})\ .
\end{equation}
Substituting the above value for $\Phi$ into eq. (\ref{kahl1}) leads to
 the  following expression:
\begin{eqnarray}
\label{kahll}
K^{(1)}&=&
 -\frac{1}{32\pi^2} \Big\{4\lambda^2 |\phi|^2
(1+\log\frac{2 M^2}{\mu^2})+2 g^2|\phi|^2\log\frac{2g^2|\phi|^2}{\mu^2}-
\nonumber\\&-&
\frac{1}{M}\Big[2\lambda^2 g (\phi+\bar{\phi})|\phi|^2
\log\frac{g^2|\phi|^2}{M^2}+\\&+&
2\lambda^2 g \big\{\bar{\phi}\big(-\frac{\phi^2}{2}+
\frac{g\hbar}{64\pi^2}\bar{D}^2
\big[\bar{\phi}(1+\log\frac{2g^2|\phi|^2}{\mu^2})\big]\big)+h.c.\big\}\Big]
\Big\}
+O(\frac{1}{M^2})\ .\nonumber
\end{eqnarray}
A substitution of    $\Phi$ from eq. (\ref{app3})
 into  the chiral potential
$W(\phi,\Phi)$  leads to the  following effective chiral potential for the light
superfield:
\begin{equation}
\label{sp}
W_{eff}=\frac{g}{3!}\phi^3
 -\frac{1}{8}\frac{\lambda^2\phi^4}{M}+
\frac{\lambda^2 g^2}{2M}
{\Big\{\frac{\hbar\bar{D}^2}{64\pi^2}
\big(\bar{\phi}
(1+\log\frac{2g^2|\phi|^2}{\mu^2})\big)\Big\}}^2
+O(\frac{1}{M^2})\ .
\end{equation}
As a result, we can now write the one-loop corrected effective action
 for light superfields,  up to the order $\frac{1}{M}$, in the  following form:
\begin{equation}
\label{eacti}
S_{eff}[\phi,\bar{\phi}]=S_{eff}^{(0)}[\phi,\bar{\phi}]+
S_{eff}^{(1)}[\phi,\bar{\phi}]\ ,
\end{equation}
where $S_{eff}^{(0)}[\phi,\bar{\phi}]$ is the  effective action at the zeroth-order
 in the expansion in
powers of the  inverse mass $M$. It takes the following form:
 \begin{eqnarray}
\label{ecl}
S_{eff}^{(0)}[\phi,\bar{\phi}]&=&
\int d^8z (\phi\bar{\phi}
 -\frac{\hbar}{32\pi^2}(4\lambda^2 |\phi|^2
(1+\log\frac{2 M^2}{\mu^2})+2 g^2|\phi|^2\log\frac{2 g^2|\phi|^2}{\mu^2}
)+\nonumber\\&+&
(\int d^6 z \frac{g}{3!}\phi^3+h.c.)\ ,
\end{eqnarray}
and $S_{eff}^{(1)}[\phi,\bar{\phi}]$ is the effective action at the
first-order in $1\over M$  expansion  and it  is of the form:
\begin{eqnarray}
\label{eqt}
S_{eff}^{(1)}[\phi,\bar{\phi}]&=&
 -\frac{\hbar}{32\pi^2 M}\int d^8 z\Big\{-2\lambda^2 g
(\phi+\bar{\phi})|\phi|^2
\log\frac{g^2|\phi|^2}{M^2}+\nonumber\\&+&
2\lambda^2 g \big\{\bar{\phi}(-\frac{\phi^2}{2}+\frac{\hbar g}{64\pi^2}
\bar{D}^2
[\bar{\phi}(1+\log\frac{2g^2|\phi|^2}{\mu^2})])+h.c.\big\}\Big\}+\nonumber\\
&+&\frac{1}{M}\Big\{\int d^6 z \Big(
\frac{1}{8}\lambda^2\phi^4-
\lambda^2 g^2
{\Big[\frac{\hbar\bar{D}^2}{64\pi^2}
\big(\bar{\phi}
\big[1+\log\frac{2g^2|\phi|^2}{\mu^2}\big]\big)\Big]}^2
\Big)+h.c.
\Big\} \ . 
\end{eqnarray}
Eqs. (\ref{eacti})-(\ref{eqt}) are our final technical  results, which are in
agreement with the decoupling theorem. 
Namely, 
the  heavy superfields at the  zeroth-order in $\frac{1}{M}$ expansion
 decouple from the effective theory;
the effective action of light superfields at the zeroth order  is
described by the Lagrangian where  all the 
effects of the heavy superfields are contained in the
redefinition (rescaling)  of  the parameters of the model, i.e.,
fields, masses,  and couplings
\cite{appcara,Collins}.  In particular, 
let  us rescale  the light 
 superfield $\phi$ in eq. (\ref{ecl}) as
$\tilde{\phi}=Z^{1/2}\phi$, where
$ Z=1-\frac{\hbar}{32\pi^2}
4\lambda^2(1+\log\frac{2 M^2}{\mu^2})\ ,
$
in order to transform the  kinetic  energy term to a canonical form.  We 
also rescale  the coupling $g$ as $\tilde{g}=Z^{-3/2}g$.
After these  transformations the effective action of light superfields
indeed takes the form
\begin{eqnarray}
\label{eac}
S_{eff}[\phi,\bar{\phi}]=\int d^8 z |\tilde{\phi}|^2
+\big[\int d^6 z\frac{1}{3!}\tilde{g}\tilde{\phi}^3+h.c.]
 -\frac{\hbar}{32\pi^2}\int d^8 z
2 \tilde{g}^2|\tilde{\phi}|^2\big(\log
\frac{2\tilde{g}^2|\tilde{\phi}|^2}{\mu^2}\big)
+O(\frac{1}{M})  ,
\end{eqnarray}
i.e., this  is the one-loop corrected effective action of the light fields
that one would have obtained within the theory  of light fields, only.
Let us also point  out that  the logarithmic term  of the form
in eq. (\ref{eac})
is due to the quantum corrections of  the self-interaction term ($\propto g$)
 of light superfields,  only; this   is
the well-known Coleman-Weinberg one-loop correction to  the effective
potential due to light superfields, only.

We  turn to the interpretation and to  determine the implications  of  the 
one-loop effective action
(\ref{eacti}) -- (\ref{eqt}) when the effective field theory is derived from a
 fundamental theory, like superstring theory where  the original 
 couplings are calculable. In this case the rescaling of the  fields and the
 couplings, performed above,  has important implications for the  quantitative
predictions for the effective   couplings at low energies.  

Let us further illustrate the significance of these effects
by imposing  the  renormalization condition $\log\frac{2 M^2}{\mu^2}=-1$
in order to cancel a term proportional to $\lambda^2$ in the zeroth-order of
$1\over
M$ expansion, i.e., this renormalization condition eliminates in the effective
action the
effect of a term responsible for the
interaction between the 
heavy and light superfields. One then arrives at the effective action
\begin{eqnarray}
S_{eff}[\phi,\bar{\phi}]&=&
\int d^8z (\phi\bar{\phi}
 -\frac{\hbar}{16\pi^2}
g^2|\phi|^2\log\frac{g^2|\phi|^2}{e M^2})
+(\int d^6 z \frac{g}{3!}\phi^3+h.c.)-\nonumber\\
&-&\frac{1}{M}\Big\{\frac{\hbar}{32\pi^2 }\int d^8 z\Big[-2\lambda^2 g
(\phi+\bar{\phi})|\phi|^2
\log\frac{g^2|\phi|^2}{M^2}+\nonumber\\&+&
2\lambda^2 g \big\{\bar{\phi}(-\frac{\phi^2}{2}+
\frac{\hbar g}{64\pi^2}\bar{D}^2
[\bar{\phi}(1+\log\frac{g^2|\phi|^2}{e M^2})])+h.c.\big\}\Big]
+\nonumber\\ &+&\Big[\int d^6 z \Big(
\frac{1}{8}\lambda^2\phi^4- \lambda^2 g^2
{\Big[\frac{\hbar\bar{D}^2}{64\pi^2}
\big(\bar{\phi}
\big[1+\log\frac{g^2|\phi|^2}{e M^2}]\big)\Big]}^2
\Big)+h.c.\Big]
\Big\}\ .
\end{eqnarray}
(Here $e=\exp(1)$).  Note that the effect of the  
 quantum  corrections   in the above action 
  grows logarithmically with $\log M$ and 
 is of the form
\begin{eqnarray}
-\frac{\hbar}{32\pi^2}
2|g\phi|^2\Big(\log\frac{2|g\phi|^2}{e M^2}\Big)
\Big\}+O(\frac{1}{M}) \ , 
\end{eqnarray}
thus, reconfirming that the   decoupling effects can be significant.

Alternatively,  we can keep $\mu$ free, and the 
the effective theory  (\ref{eacti}) -- (\ref{eqt}) 
can be viewed as  specified at  low  energies $\mu\ll M
$. Again,  the obtained
effective action (\ref{ecl}) has quantum corrections that 
are  proportional to $\log\frac{2 M^2}{\mu^2}$, i.e., these
 corrections grow as  $\log {M\over \mu}$ when
 ${M\over \mu}\to\infty$. 

Within superstring theory 
the above corrections can now have sizable
 calculable predictions for the low energy couplings.
Let us illustrate the effect 
within a class of perturbative heterotic
 string models with an anomalous
$U(1)$; after the vacuum restabilization there are in  general
renormalizable  interactions between the light and heavy fields, with
the mass of order
 $M\sim
 M_{String}\sim
10^{17}$ GeV .
 The one-loop decoupling effects
can significantly change
  the low energy predictions for the couplings at 
the electro-weak scale
 ($ \mu\sim 1$ TeV). For example, in a class of string models discussed in,
 \cite{cceel2}--\cite{Cv4},  typical values of the
  couplings, calculated at $M_{String}$, are
$\lambda =g=g_{gauge}$. Here
$g_{gauge}\sim 0.8 $ is the value of the gauge coupling at
at $M_{String}$.  Renormalization group equations then determine
the values of $\lambda$ and $g$ couplings at low energies $\mu$. 
(For fields charged under the non-Abelian gauge group factors 
such large tri-linear couplings are driven to the infrared fixed points
governed by the infrared values of the non-Abelian gauge couplings.) 
Due to the  above one-loop decoupling effects there is now
also an additional correction to the effective tri-linear coupling
$g$ for the light fields; it  is of the order
 of  ${{\lambda (\mu )^2}\over {16\pi^2}}\log {{M^2}\over{\mu^2}}\sim 0.26$. 
 (We used
the typical values $M\sim M_{String}\sim 10^{17}$ GeV, $\mu\sim 1$ TeV and
$\lambda(\mu)\sim
0.8$.) This specific example
   illustrates  that  due to  the one-loop
 decoupling effects,   the actual predictions 
  for  the tri-linear couplings at low energies  could be
 corrected by
  $10-50\%$!

\vspace*{1mm}

{\bf Acknowledgements.} M.C. would like to thank P.
Langacker for discussions and L. Everett for comments. The work by I.L.B. and
A.Yu.P.
was supported
in part by INTAS grant, INTAS-96-0308; RFBR grant, project
No 99-02-16617; RFBR-DFG grant, project No 96-02-00180; GRACENAS, project
No 97-6.2-34.  The work of M.C.\ was supported in part by U.S. Department of
Energy Grant No.  DOE-EY-76-02-3071. I.L.B. and M.C. would like to thank the
organizers of the 32nd
International Symposium Ahrenshoop on the Theory of Elementary Particles,
where the work was initiated,  for
hospitality.


\begin{thebibliography}{100}
\bibitem{appcara}
K. Symanzik, Comm. Math. Phys. {\bf 34}, 7 (1973);
T. Appelquist and J. Carrazone, Phys. Rev. {\bf D11}, 2856 (1975).
\bibitem{Collins}
E. Witten, Nucl.Phys. {\bf B104},  445 (1976); J.C. Collins,
F. Wilczek and A. Zee, Phys. Rev. {\bf D18},  242 (1978); B.A. Ovrut and
H.T. Schitzer,
Phys. Rev. {\bf D21} 3369 (1980).
J.C. Collins, Renormalization   (Gordon and Breach, N.Y.), 1982.
\bibitem{Shenkeretal}
D. Friedan, E. Martinec and S. Shenker,
Nucl. Phys. {\bf B271}, 93 (1986).
\bibitem{bailin} L.~Dixon, E.~Martinec, D.~Friedan and S.~Shenker,
Nucl. Phys. {\bf B282}, 13 (1987);
M. Cveti\v c, Phys. Rev. Lett. {\bf 59}, 2829 (1987); D. Bailin  and  A. Love,
 Phys. Lett. {\bf B260}, 56 (1991);
 S. Kalara, Jorge L. Lopez and  D.V. Nanopoulos, Nucl. Phys. {\bf B353},
 650 (1991).
\bibitem{GSW} M.B. Green, J.H.
Schwarz, and E. Witten, Superstring Theory (Cambridge Univ.  Press, 1987),
Vol.2.
\bibitem{dsw}{M.~Dine, N.~Seiberg and E.~Witten, Nucl. Phys. {\bf B289}, {585}
(1986); L. Dixon
and V. Kaplunovsky, unpublished.}
\bibitem{atick}{J.~Atick, L.~Dixon and A.~Sen,  Nucl. Phys. {\bf B292}, {109}
(1987);
M.~Dine, I.~Ichinose and N.~Seiberg,  Nucl. Phys. {\bf B293},{253} (1987);
M.~Dine and C.~Lee,  Nucl. Phys. {\bf B336},{317} (1990).}\
\bibitem{CELW}
M. Cveti\v c, L. Everett, P. Lagacker and J. Wang, Blowing up Type I Z3
Orientifold, {\it Preprint}
hep-th/9903051.
 \bibitem{cceel2} G. Cleaver, M. Cveti\v{c}, J.R.
Espinosa, L. Everett, and P. Langacker,  Nucl. Phys. {\bf B525}, 3 (1998);
 G. Cleaver, M. Cveti\v{c}, J.R. Espinosa,
L. Everett and P. Langacker,
Flat direction in three generation free fermionic string models,
{\it Preprint} hep-th/9805133, to be published in
Nucl. Phys. {\bf B}.
\bibitem{Cv4}
G. Cleaver, M. Cveti\v{c}, J.R. Espinosa, L. Everett,  and P. Langacker,
Phys.  Rev. {\bf D59}, 55 (1999); Physics Implications of Flat
Directions
in Free Fermionic Superstring Models II: Renormalization Group
Analysis, {\it Preprint} hep-ph/9811355, to be published in Phys. Rev.
{\bf D}.

\bibitem{Burgess} C.P. Burgess, A. Font and
F. Quevedo,  Nucl. Phys. {\bf B272},  661 (1986);
A. Font  and F. Quevedo, Phys. Lett. {\bf B184}, 45 (1987).

\bibitem{katehou} E. Katehou and G.G. Ross, Nucl. Phys. {\bf B299}, 484
(1988).
\bibitem{cew}
M. Cveti\v{c}, L. Everett  and J. Wang. Nucl. Phys. {\bf B538}, 52 (1999).

\bibitem{dine} N. Arkani-Hamed, M. Dine and S. Martin,
Dynamical supersymmetry breaking in models with a Green-Schwarz mechanism,
{\it Preprint} hep-ph/9803432.
\bibitem{Buchbinderetal}  I.L. Buchbinder, M. Cveti\v c, and
A.Yu. Petrov, in preparation.
\bibitem{BK0} I.L. Buchbinder and  S.M. Kuzenko, Ideas and Methods
of Supersymmetry and Supergravity or a Walk Through Superspace (IOP
Publishing, Bristol and Philadelphia, 1995),  (revised edition, 1998).
\bibitem{West2} P. West, Phys. Lett. {\bf B261}, 396 (1991);
I. Jack, D.R.T. Jones and  P. West, Phys. Lett. {\bf B258}, 382 (1991); I.L.
 Buchbinder, S.M. Kuzenko and  A.Yu. Petrov, Phys. Lett. {\bf B321}, 372
(1994).
 \bibitem{Buch1}
 I.L. Buchbinder, S.M. Kuzenko and J.V. Yarevskaya,
Yad.Fiz. (Physics of Atomic Nuclei, in Russian) {\bf 56}, 202
(1993); Nucl. Phys. {\bf B411}, 665  (1994);
 I.L. Buchbinder, S.M. Kuzenko and A.Yu. Petrov,
Yad.Fiz. (Physics of Atomic Nuclei, in Russian) {\bf 59}, 157
(1996).

\end{thebibliography}
\end{document}